\documentclass{PoS}
\usepackage{cite}
\usepackage[utf8]{inputenc}
\usepackage{amsmath}
\usepackage{amsfonts}
\usepackage{amssymb}
\usepackage{amsthm}
\usepackage{enumitem}
\usepackage[T1]{fontenc}
\usepackage{dcolumn}
\usepackage{bm}
\usepackage{bbm}
\usepackage{cancel}
\usepackage[ngerman,english]{babel}
\usepackage[pdftex]{graphicx}
\usepackage{tikz}
\usetikzlibrary{arrows}
\usepackage{tabularx}
\usepackage{booktabs}
\usepackage{tensor}
\usepackage{xcolor}
\usepackage{mathtools}
\usepackage{float}
\usepackage{sidecap}
\usepackage{relsize}
\usepackage{exscale}
\usepackage{textpos}
\definecolor{darkblue}{rgb}{0,0,.5}
\usepackage{times}

\newcommand{\e}{\operatorname{e}}

\newcommand{\SO}[1]{\operatorname{SO}\left(#1\right)}

\newcommand{\Sph}[1]{\operatorname{S}^{#1}}
\newcommand{\of}[1]{\mathop{\left(#1\right)}}
\newcommand{\fof}[1]{\mathop{\left[#1\right]}}
\newcommand{\cof}[1]{\mathop{\left\{#1\right\}}}

\newcommand{\avof}[1]{\mathop{\left\langle #1\right\rangle}}

\newcommand{\eqb}{\begin{equation}}
\newcommand{\eqe}{\end{equation}}

\newcommand{\idd}[2]{\mathrm{d}^{#2}\,#1}

\newcommand{\DD}[1]{\mathcal{D}\left[#1\right]}

\newcommand{\partd}[2]{\frac{\partial #1}{\partial #2}}

\newcommand{\ford}{1^{\text{st}}\,\text{order}}
\newcommand{\sord}{2^{\text{nd}}\,\text{order}}

\newcommand{\nord}[1]{#1^{\text{th}}\,\text{order}}

\newcommand{\abs}[1]{\left|#1\right|}
\newcommand{\op}[1]{\operatorname{#1}}

\renewcommand*\[{\begin{equation}}
\renewcommand*\]{\end{equation}}

\let\oldstackrel\stackrel
\renewcommand*\stackrel[2]{{\scriptstyle\oldstackrel{#1}{#2}}}

\definecolor{emphcol}{RGB}{0,0,0}
\let\oldemph\emph
\renewcommand*\emph[1]{\oldemph{\textcolor{emphcol}{#1}}}

\title{
\begin{textblock*}{100pt}(316pt,-154pt)
\textnormal{\small \texttt{CERN-PH-TH/2013-261}}
\end{textblock*}
Euclidean Dynamical Triangulation revisited:\\
is the phase transition really first order?}

\ShortTitle{EDT revisited: is the phase transition really first order?}

\author{\speaker{Tobias Rindlisbacher}\\
        Institute for Theoretical Physics, ETH Z\"urich, CH - 8093 Z\"urich, Switzerland\\
        E-mail: \email{rindlisbacher@itp.phys.ethz.ch}}

\author{Philippe de Forcrand\\
        Institute for Theoretical Physics, ETH Z\"urich, CH - 8093 Z\"urich, Switzerland and\\
	CERN, Physics Department, TH Unit, CH-1211 Gen\`eve 23, Switzerland\\
        E-mail: \email{forcrand@itp.phys.ethz.ch}}

\abstract{The transition between the two phases of 4D Euclidean Dynamical Triangulation \cite{Ambjorn} was long believed to be of second order until in 1996 first order behavior was found for
sufficiently large systems \cite{Bialas,deBakker}. However, one may wonder if this finding was affected by the numerical methods used: to control volume fluctuations, in both studies
\cite{Bialas,deBakker} an artificial harmonic potential was added to the action; in \cite{deBakker} measurements were taken after a fixed number of \emph{accepted} instead of \emph{attempted} moves
which introduces an additional error. Finally the simulations suffer from strong critical slowing down which may have been underestimated.\\
In the present work, we address the above weaknesses: we allow the volume to fluctuate freely within a fixed interval; we take measurements after a fixed number of attempted moves; and we overcome critical
slowing down by using an optimized parallel tempering algorithm \cite{Bauer}. With these improved methods, on systems of size up to $N_{4}=64$k 4-simplices, we confirm that the phase transition is
first order.}
        
\FullConference{31st International Symposium on Lattice Field Theory LATTICE 2013\\
		 July 29 - August 3, 2013\\
		 Mainz, Germany}

\begin{document}

\section{Introduction}\label{sec:intro}
The transition between the two phases of 4D Euclidean Dynamical Triangulation (EDT)\cite{Ambjorn} was first believed to be of $\sord$ until 1996 where in \cite{Bialas} for the first time $\ford$
behavior was reported for a system consisting of $N_{4}=32$k 4-simplices. Shortly afterwards this finding was verified in \cite{deBakker} and extended to larger systems with $N_{4}=64$k. However, we
were not completely convinced by the numerical methods used in the latter work; there were three things which disturbed us:
\vspace{-7pt}
\begin{enumerate}\itemsep -2pt
\item Measurements were taken after a fixed number of \textbf{accepted} (instead of attempted) moves, which introduces a systematic error.
\item The use of an artificial harmonic potential to control volume fluctuations also introduces a systematic error.
\item Autocorrelation and thermalization time could easily have been underestimated.
\end{enumerate}
\vspace{-5pt}
Therefore we wanted to crosscheck these old results with our own, hopefully correct code which satisfies detailed balance, uses a potential well instead of a harmonic potential to control
volume fluctuations, and makes use of \emph{parallel tempering} to cope with critical slowing down.\\

\subsection{The EDT Model}
In 4-dimensional Euclidean Dynamical Triangulation (EDT)\cite{Ambjorn} the formal path integral for \emph{Euclidean} (local $\SO{4}$ instead of $\SO{3,1}$ symmetry) \emph{gravity},
\[
Z\,=\,\int\DD{g\indices{_\mu_\nu}} \e^{-\,S_{EH}\fof{g\indices{_\mu_\nu}}},\label{eq:epf}
\]
with $S_{EH}=-\frac{1}{16 \pi G}\,\int\idd{x}{4}\sqrt{g}\of{R-2\Lambda}$ being the \emph{Einstein-Hilbert action}, is regularized by approximating the configuration space (space of all
diffeomorphism inequivalent 4-metrics) with the space of simplicial piecewise linear (PL) manifolds consisting of equilateral 4-simplices with fixed edge length $a$ (such manifolds are also called
\emph{abstract triangulations}). Under such a discretization, the Einstein-Hilbert action turns into the \emph{Einstein-Regge action} which for equilateral 4-simplices and a space-time of topology
$\Sph{4}$ takes the simple form $S_{ER}=-\kappa_{2}N_{2}+\kappa_{4}N_{4}$, where the $N_{i}$ label the number of $i$-simplices in the PL manifold, $\kappa_{2}=\frac{V_{2}}{4\,G}$ and
$\kappa_{4}=\frac{10 \op{arccos}\of{1/4} V_{2}+\Lambda V_{4}}{8 \pi G}$, with $V_{n}=a^{n}\frac{\sqrt{n+1}}{n! \sqrt{2^{n}}}$ being the volume of a $n$-simplex and $G$, $\Lambda$ the
gravitational and cosmological coupling respectively. The partition function \eqref{eq:epf} can now be written as
\[
Z\of{\kappa_{2},\kappa_{4}}\,=\,\sum\limits_{T}\,\frac{1}{C_{T}}\,\e^{\kappa_{2}\,N_{2}\of{T}\,-\,\kappa_{4}\,N_{4}\of{T}}\,=\,\sum\limits_{N_{4}}\,Z\of{\kappa_{2},N_{4}}\,\e^{-\kappa_{4}\,N_{4}},\label{eq:edtpf}
\]
where the sum after the first equality sign runs over all abstract triangulations $T$ of $\Sph{4}$ and $C_{T}$ is for each $T$ a corresponding symmetry factor\footnote{$C_{T}$ is assumed to be
$\sim\,1$ for sufficiently large systems.}. The second equality sign defines the canonical partition function $Z\of{\kappa_{2},N_{4}}$. The partition function \eqref{eq:edtpf} is suitable for use in a
Markov chain Monte Carlo simulation with Metropolis updates consisting of the so-called \emph{Pachner moves} \cite{Ambjorn}.  

\subsection{Phase Diagram}
The grand canonical partition function \eqref{eq:edtpf} is finite only if $\kappa_{4}>\kappa_{4}^{cr}\of{\kappa_{2}}$. We therefore have a critical line for convergence in the
$\of{\kappa_{2},\kappa_{4}}$-plane, given by $\kappa_{4}^{cr}\of{\kappa_{2}}$. To obtain the thermodynamic limit $\of{N_{4}\rightarrow\infty}$ we have to ensure that 
$\kappa_{4}\overset{\scriptscriptstyle N_{4}\rightarrow\infty}{\longrightarrow}\kappa_{4}^{cr}\of{\kappa_{2}}$.
For quasi-canonical simulations around some fixed volume $\bar{N}_{4}$, it follows from \eqref{eq:edtpf} that we can define a pseudo-critical
$\kappa_{4}^{pcr}\of{\kappa_{2},\bar{N}_{4}}=\partd{\ln\of{Z\of{\kappa_{2},N_{4}}}}{N_{4}}\big|_{\scriptscriptstyle N_{4}=\bar{\scriptscriptstyle N}_{4}}$, which corresponds to the value of $\kappa_{4}$ for which the
$N_{4}$-distribution is flat around $\bar{N}_{4}$.\\
\begin{figure}[htbp]
\centering
\begin{minipage}[t]{0.49\linewidth}
\centering
\includegraphics[width=\linewidth]{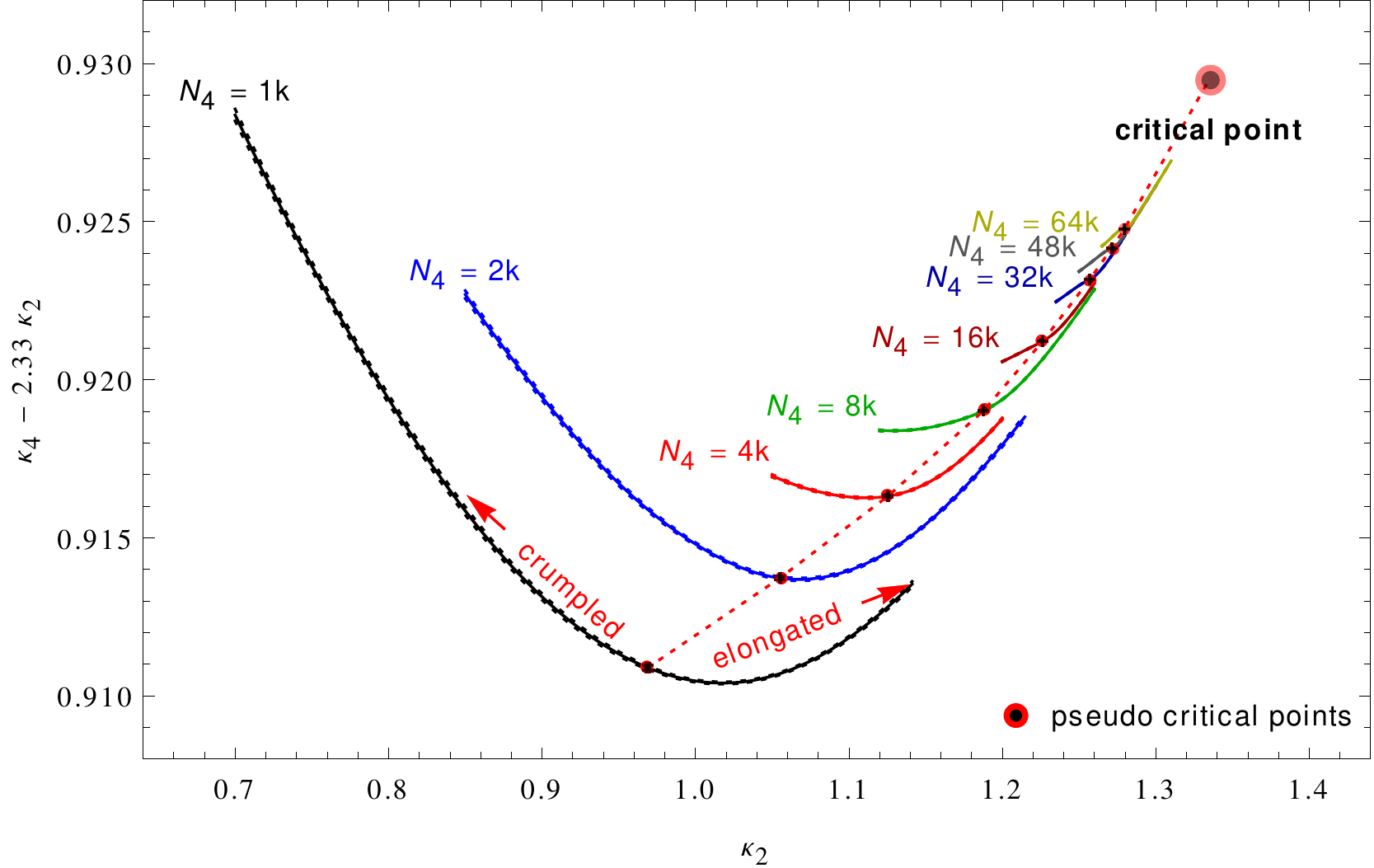}
\caption{Phase diagram for 4D EDT. The Figure shows $\kappa_{4}^{pcr}\of{\kappa_{2},N_{4}}$ as a function of $\kappa_{2}$ for different $N_{4}$ together with the
corresponding pseudo-critical points $\of{\kappa_{2}^{pcr}\of{N_{4}},\kappa_{4}^{pcr}\of{\kappa_{2}^{pcr}\of{N_{4}},N_{4}}}$. The dotted red line separates the \emph{crumpled} from the
\emph{elongated} phase; in the limit $N_{4}\,\rightarrow\,\infty$ this line ends at the \emph{critical point}: $\of{\kappa_{2}^{cr},\kappa_{4}^{cr}}$. To improve readability, the
y-axis shows $\of{\kappa_{4}-2.33\,\kappa_{2}}$ instead of $\kappa_{4}$ itself.}
  \label{fig:edtpd}
\end{minipage}\hfill
\begin{minipage}[t]{0.49\linewidth}
\centering
\includegraphics[width=0.97\linewidth]{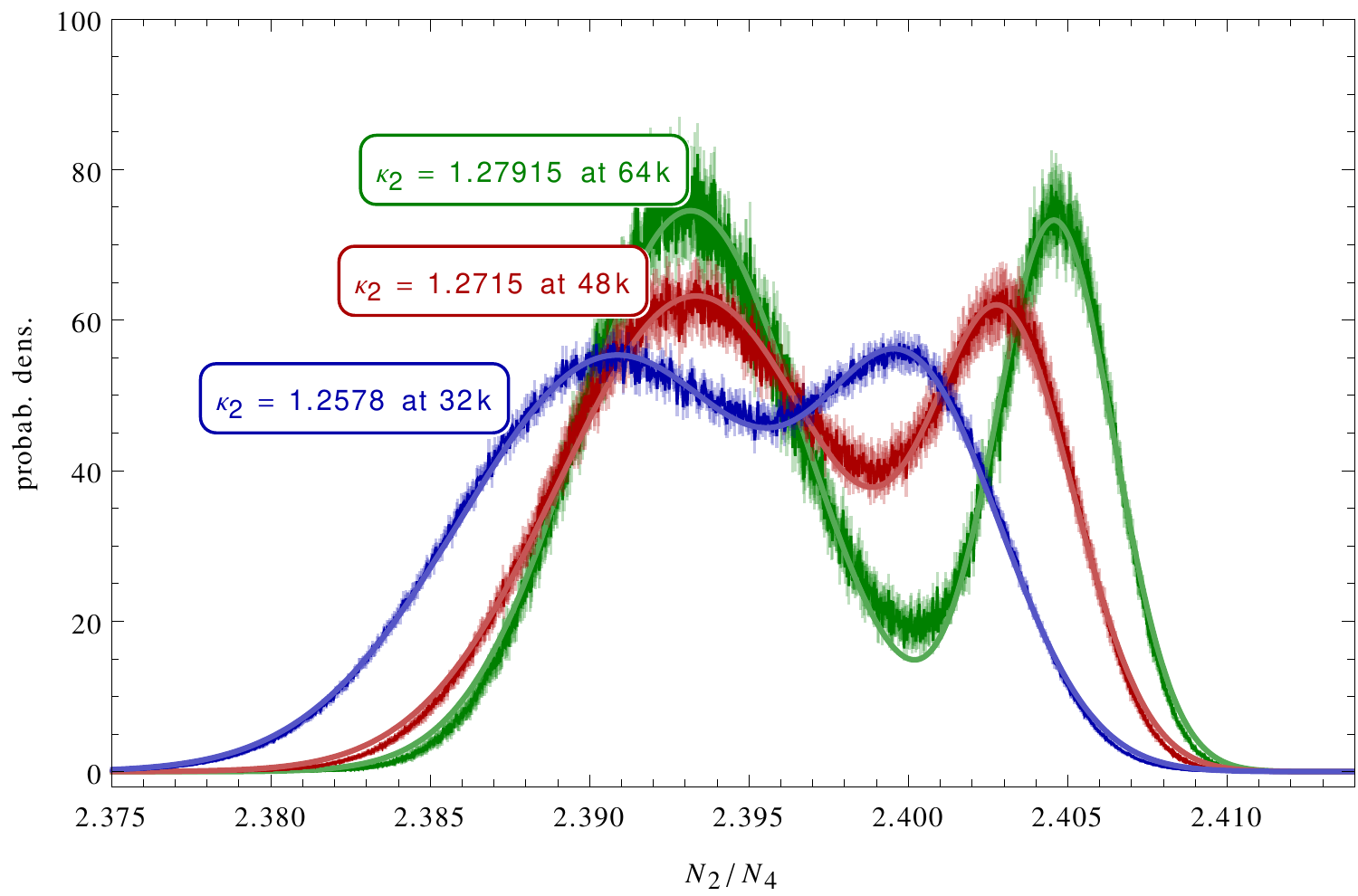}
\caption{$N_{2}$-distribution for systems of size $N_{4}\,=\,$ 32k (blue), 48k (red) and 64k (green): the solid lines are double-Gaussian fits to the data. It can be seen that the
double peak structure becomes more pronounced with increasing system size and there is no sign that the peaks will merge again in the
thermodynamic limit. This is characteristic of a $\ford$ transition.}
\label{fig:N2dist}
\end{minipage}
\end{figure}

\begin{figure}[htbp]
\centering
\begin{minipage}[t]{0.49\linewidth}
\centering
\includegraphics[width=0.95\linewidth]{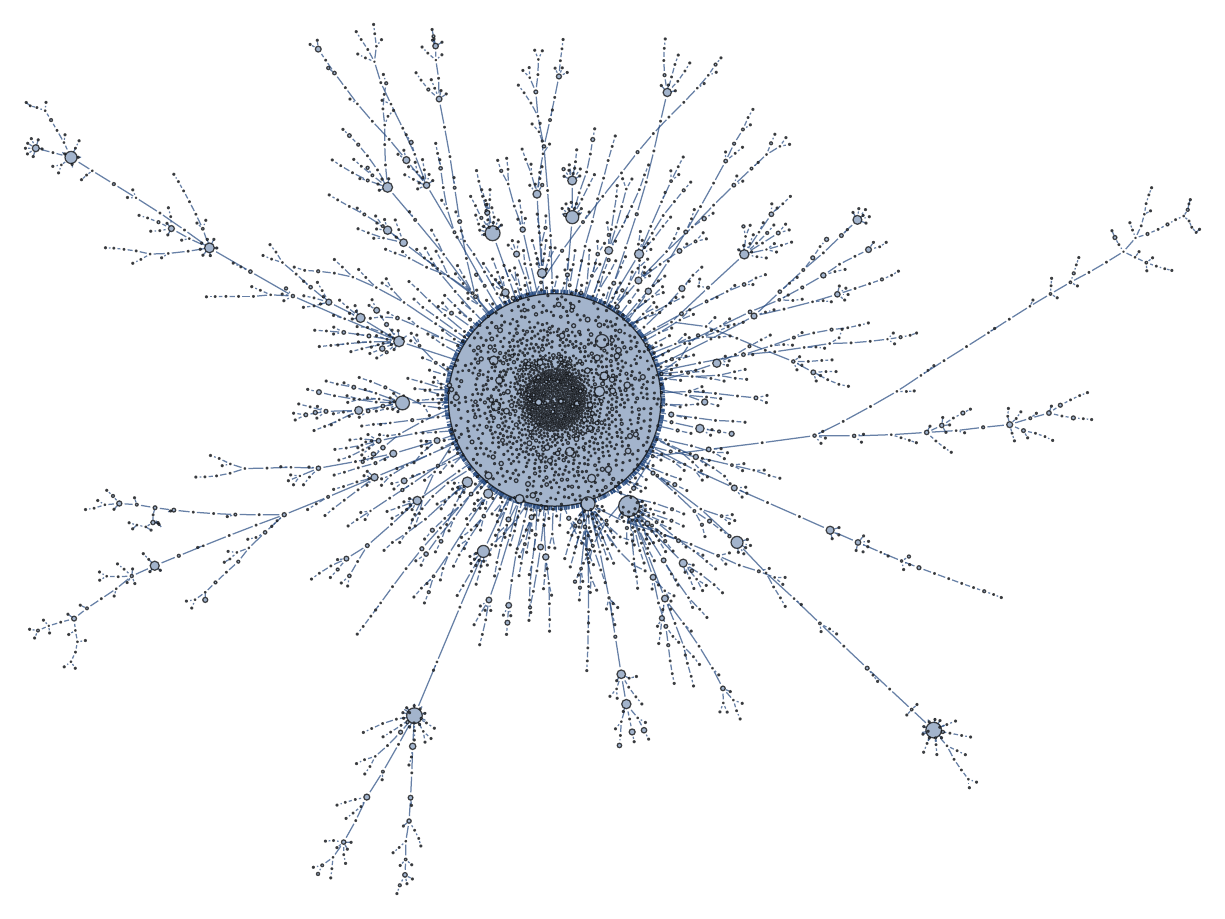}
\end{minipage}\hfill
\begin{minipage}[t]{0.49\linewidth}
\centering
\includegraphics[width=0.83\linewidth]{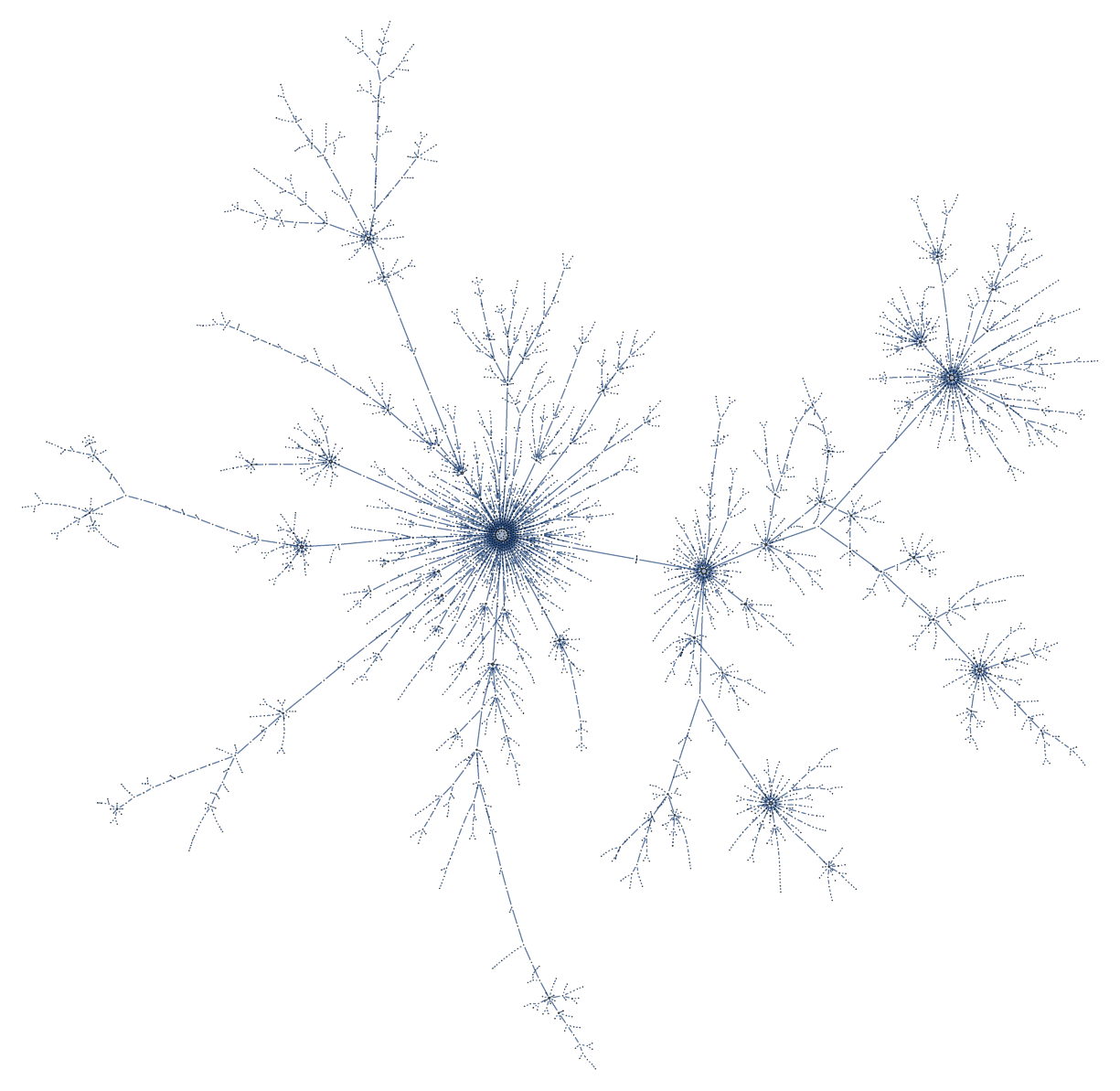}
\end{minipage}
\caption{Representative configurations in the crumpled (left, $\kappa_{2}=1.24$) and elongated (right, $\kappa_{2}=1.27$) phase at system size $N_{4}\approx 32$k: in the
crumpled phase, the triangulation consists of one large, highly connected bunch with outgrowths which are at least an order of magnitude smaller. In the elongated phase on the other hand, although a
largest component still exists and may be called ``mother universe'', it is much smaller than in the crumpled phase and some of its outgrowths (the ``babies'') are of comparable size.}
\label{fig:butrees}
\end{figure}

For constant $N_{4}$, we can define (see Fig. \ref{fig:edtpd}) a line $\kappa_{4}^{pcr}\of{\kappa_{2},N_{4}}$ as a function of $\kappa_{2}$, along which two phases are separated by a pseudo-critical
point at $\kappa_{2}=\kappa_{2}^{pcr}\of{N_{4}}$. For $\kappa_{2}<\kappa_{2}^{pcr}\of{N_{4}}$ we are in the \emph{crumpled phase} where a typical configuration is
highly collapsed in the sense that the distance between any two 4-simplices is very short, leading to a large (infinite) Hausdorff dimension. For $\kappa_{2}>\kappa_{2}^{pcr}\of{N_{4}}$ we are
in the \emph{elongated phase} with Hausdorff dimension $\sim\,2$, where a typical configuration consists of a so-called \emph{baby-universe tree}: the total volume is subdivided into smaller parts,
the \emph{baby-universes}, which are pairwise connected by only a small neck. This structure is hierarchical in a treelike manner: consider the largest baby-universe as ``mother'' with outgrowing
smaller ``babies'' which in turn give birth to their own ``babies'', and so on (see Fig. \ref{fig:butrees}).\\
The true \emph{critical point} in the thermodynamic limit is obtained as
\[
\of{\kappa_{2}^{cr},\kappa_{4}^{cr}}\,=\,\lim\limits_{N_{4}\rightarrow\infty}\,\of{\kappa_{2}^{pcr}\of{N_{4}},\kappa_{4}^{pcr}\of{\kappa_{2}^{pcr}\of{N_{4}},N_{4}}}.
\]

\section{Simulation Methods}\label{sec:simulationmethods}
We are interested in the canonical system, but as there is no set of ergodic moves known for the space of triangulations of fixed volume, we can only run \emph{quasi-canonical}\footnote{I.e. grand
canonical simulations where the volume is constrained to fluctuate around some desired volume $\bar{N}_{4}$.} Markov chain Monte Carlo simulations with local updates consisting of the five Pachner
moves.  

\subsection{Detailed Balance}\label{sec:detailedbalance}
Calling $T_{k}$ the current triangulation in our Markov chain, we obtain $T_{k+1}$ as follows: 
\vspace{-5pt}
\begin{enumerate}\itemsep -2pt
 \item randomly choose a move type $n\in\cof{0,\ldots,4}$
 \item randomly choose one of the $N_{4}$ $4$-simplices of $T_{k}$ and call it $D$
 \item randomly choose one of the $\binom{5}{n+1}$ $n$-simplices contained in $D$ and call it $S$
 \item\label{it:acctest} perform a Metropolis test with acceptance probability $p_{n}\of{T_{k},S}$:
\vspace{-5pt}
 \begin{itemize}\itemsep -2pt
  \item accept: $T_{k+1}$ is obtained from $T_{k}$ by applying the $n$-move at $S$
  \item reject: $T_{k+1} = T_{k}$
 \end{itemize}
\vspace{-5pt}
\end{enumerate}
\vspace{-4pt}
The acceptance probability at step \ref{it:acctest} is given by \cite{deBakker2}
\[
p_{n}\of{T,S}=
\begin{cases}
p_{n}\of{N_{4}\of{T}}\quad &\text{\small
if}\quad\text{\small $n$-move possible at $S\,\in\,T$}\\
0\quad &\text{\small else}
\end{cases},\label{eq:transprobab}
\]
where $p_{n}\of{N_{4}}=\textstyle\min\cof{1,\frac{\scriptstyle N_{4}}{\scriptstyle N_{4}+\Delta N_{4}\of{n}}\,\e^{\kappa_{2}\Delta N_{2}\of{n}-\kappa_{4}\Delta N_{4}\of{n}}}$ is the
so-called \emph{reduced transition probability}, $\Delta N_{i}\of{n}$ labels the change of $N_{i}$ under a $n$-move, and a $n$-move is considered as possible at $S$ if $S$ is contained in
$\of{5-n}$ $4$-simplices\footnote{If a $n$-simplex is contained in $\of{5-n}$ 4-simplices, the complex spanned by these 4-simplices can be replaced by another complex which has the same boundary but
consists of $\of{n+1}$ 4-simplices which share a common $\of{d-n}$-simplex (orthogonal/dual to the initial $n$-simplex). This is precisely the action of a Pachner $n$-move.} and the application of the
move does not violate the \emph{manifold constraint} which requires that the move will not produce any simplex which already exists in the triangulation.\\

\subsection{Controlling the Volume}
In previous work \cite{Ambjorn,Bialas,deBakker}, the volume was controlled by adding a harmonic potential, $U=\frac{\delta}{2}\of{N_{4}-\bar{N}_{4}}^{2}$
to the Einstein-Regge action. This of course introduces a systematic error for all moves which change $N_{4}$. We therefore decided to rather use an infinite potential well of some reasonable width
$w\approx 2\sigma\of{N_{2}}/2.5$, where $2.5=\max\limits_{n}\cof{\frac{\Delta N_{2}\of{n}}{\Delta N_{4}\of{n}}}$ and $\sigma\of{N_{2}}$ is the square root of the $N_{2}$-susceptibility.\\
With such a potential well we can not use the saddle point expansion method from \cite{deBakker2} to tune $\kappa_{4}$ to its pseudo-critical value $\kappa_{4}^{pcr}\of{\kappa_{2},\bar{N}_{4}}$.
Instead we made use of a method mentioned in \cite{Bruegmann}: as the $N_{4}$-histogram has to be flat around $\bar{N}_{4}$ if $\kappa_{4}=\kappa_{4}^{pcr}\of{\kappa_{2},\bar{N}_{4}}$, we have
that
\[
\textstyle\bar{p}_{4}^{geo}\of{\bar{N}_{4}}\,p_{4}^{pcr}\of{\bar{N}_{4}}\,=\,\bar{p}_{0}^{geo}\of{\bar{N}_{4}+\Delta N_{4}\of{4}}\,p_{0}^{pcr}\of{\bar{N}_{4}+\Delta
N_{4}\of{4}},\label{eq:totmoveequil}
\]
where $p_{n}^{pcr}\of{N_{4}}$ is the reduced transition probability $p_{n}\of{N_{4}}$ from Sec. \ref{sec:detailedbalance} with $\kappa_{4}=\kappa_{4}^{pcr}\of{\kappa_{2},\bar{N}_{4}}$ and
$\bar{p}_{n}^{geo}\of{N_{4}}$ is the average probability to select within a configuration of size $N_{4}$ a $n$-simplex where a $n$-move can be applied (we call these \emph{geometric probabilities}).
One can then solve for $\kappa_{4}^{pcr}\of{\kappa_{2},\bar{N}_{4}}$ which leads to
\[
\kappa_{4}^{pcr}\of{\kappa_{2},\bar{N}_{4}}\,=\,\frac{1}{\Delta N_{4}\of{4}}\fof{\ln\of{\frac{\bar{p}_{4}^{geo}\of{\bar{N}_{4}}}{\bar{p}_{0}^{geo}\of{\bar{N}_{4}+\Delta
N_{4}\of{4}}}}\,-\,\ln\of{1+\frac{\Delta N_{4}\of{4}}{\bar{N}_{4}}}}\,+\,\frac{\Delta N_{2}\of{4}}{\Delta N_{4}\of{4}}\,\kappa_{2}.\label{eq:geok4pcr}
\]
As the 4-move is always possible, only $\bar{p}_{0}^{geo}$ has to be measured.

\subsection{Autocorrelation Time}
Critical slowing down is much worse for $\ford$ transitions than for $\sord$. This is because transitions between the two phases are \textbf{exponentially} suppressed with increasing system size. To
overcome this problem, we use an optimized \emph{parallel tempering} algorithm as described in \cite{Bauer}: for a fixed average volume $\bar{N}_{4}$, we simulate in parallel 48 systems (called
\emph{replicas}) along the pseudo-critical line $\kappa_{4}^{pcr}\of{\kappa_{2},\bar{N}_{4}}$, such that they connect a region with fast relaxation in the crumpled phase with a region with fast
relaxation in the elongated phase and thereby pass through the critical region around $\kappa_{2}^{pcr}\of{\bar{N}_{4}}$. At regular intervals, a swap of configurations between neighboring ensembles
is attempted. This motion of configurations through coupling space permits a faster evolution. We start with equally spaced $\kappa_{2}$ values and apply after some runtime
the optimization procedure of \cite{Bauer} which gives us a new set of couplings for which the configuration exchange between replicas is more frequent. A study of the efficiency of this procedure
can be found in \cite{Bauer}. 

\section{Data Analysis and Results}\label{sec:results}
Due to the use of a potential well instead of a harmonic potential to control the system volume and due to the tuning of $\kappa_{4}$ to its pseudo-critical value, we have significant volume
fluctuations in the data which also affect for example the $N_{2}$ distribution. To take this into account, we project the data in the $\of{N_{2},N_{4}}$-plane along the "correlation direction" before
evaluating any observables, i.e. instead of $N_{2}$ we use
\[
\bar{N}_{2}\,=\,N_{2}\,-\,\frac{\avof{\of{N_{2}-\avof{N_{2}}}\of{N_{4}-\avof{N_{4}}}}}{\avof{N_{4}-\avof{N_{4}}}^{2}}\,\of{N_{4}\,-\,\avof{N_{4}}}\label{eq:n2bar}
\]
to evaluate observables depending on $N_{2}$. We checked that this leads to the same results as when evaluating the observables only on data subsets corresponding to single, fixed $N_{4}$ values. From
now on, when talking about the $N_{2}$-distribution, we mean the corrected, fixed $N_{4}$-version \eqref{eq:n2bar}.\\
After that, we use \emph{multi-histogram reweighting} \cite{Ferrenberg} with respect to $\kappa_{2}$\footnote{The parallel tempering optimization procedure mentioned above also leads to a good
distribution of simulation points for the reweighting.}. The errors are determined with the Jack-Knife method with 20 sets. In multi-histogram reweighting, these sets consist of the simultaneous data
of all the simulations at different $\kappa_{2}$ values, therefore cross-correlations should automatically be taken into account. 

\subsection{$N_{2}$ Distribution}
As stated in \cite{deBakker}, the $N_{2}$-distribution starts to be double-peaked for systems consisting of more than $\sim 32000$ 4-simplices. The distributions for systems with 32k, 48k and 64k
4-simplices are shown in Figure \ref{fig:N2dist}. It can be seen that the double peak structure becomes more pronounced with increasing system size and that there is so far no sign that the peaks will
merge again in the thermodynamic limit. This behavior is characteristic of a $\ford$ transition. 

\subsection{Scaling of $B_{4}$}
\begin{figure}[htbp]
\begin{minipage}[t]{0.49\linewidth}
\centering
\includegraphics[width=\linewidth]{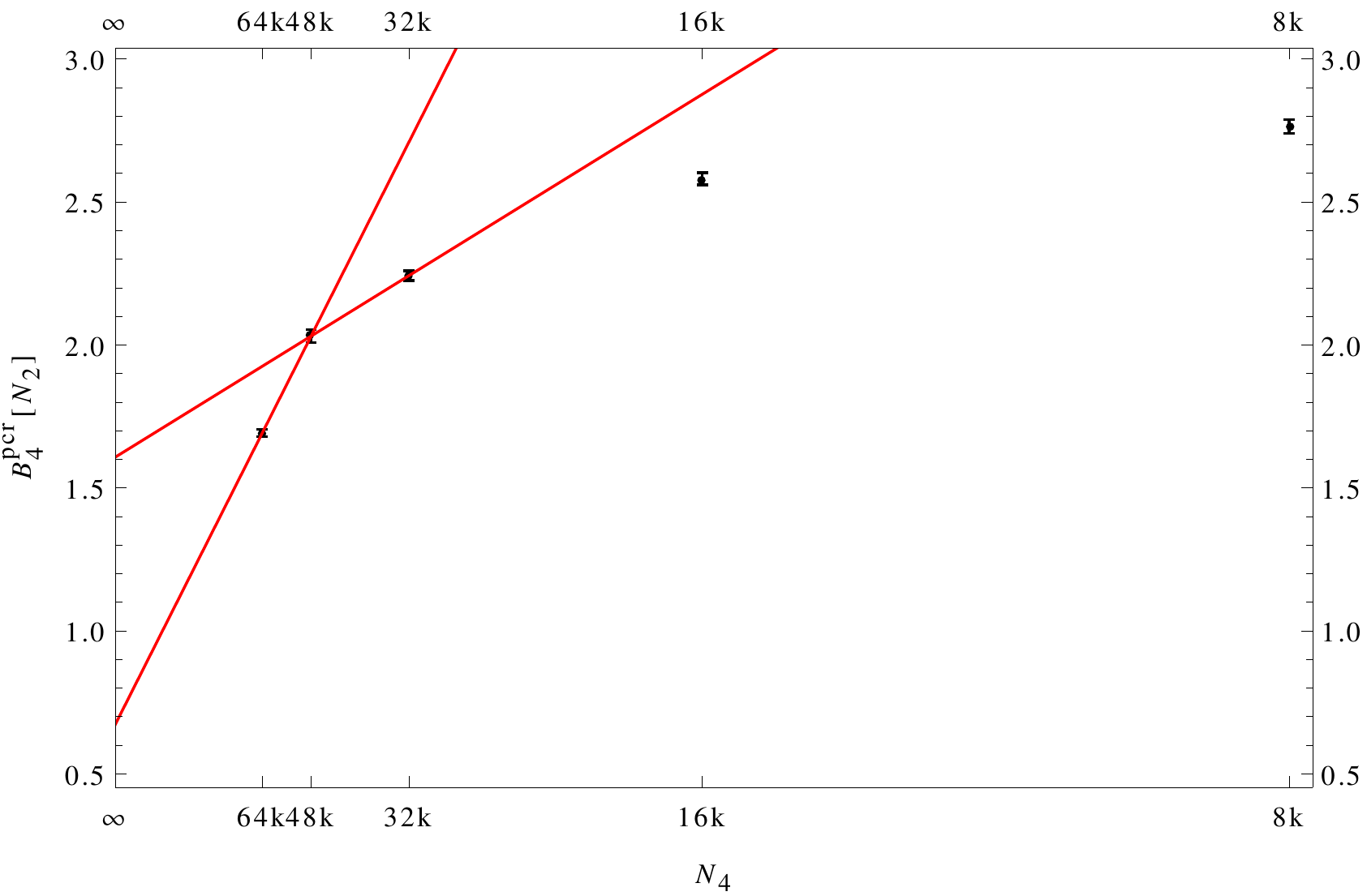}
\caption{Binder cumulant $B_{4}^{pcr}\fof{N_{2}}$ as a function of $1/N_{4}$. The red lines correspond to fits of the form \eqref{eq:b4fss} with $\omega=1$ ($\ford$ transition) to the
data of the largest and second largest pair of systems. The value $B_{4}^{cr}\fof{N_{2}}\approx 0.7$ obtained from the largest pair is too small, whereas the fit for the second largest pair yields a
too large value: $B_{4}^{cr}\fof{N_{2}}\approx 1.6$.}
  \label{fig:B4scale1}
\end{minipage}\hfill
\begin{minipage}[t]{0.49\linewidth}
\centering
\includegraphics[width=0.965\linewidth]{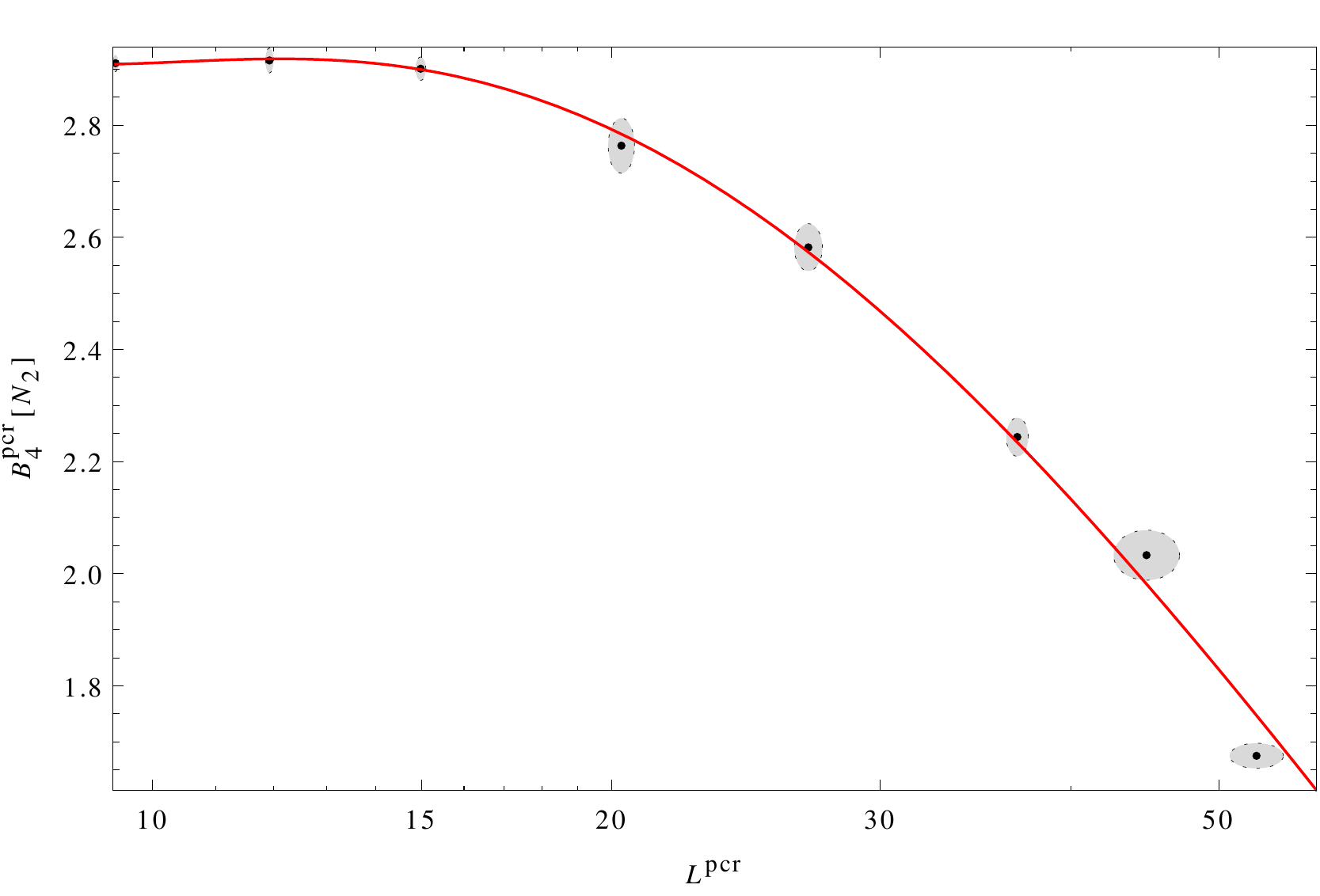}
\caption{Binder cumulant $B_{4}^{pcr}\fof{N_{2}}$ as a function of average linear system size $L^{pcr}\of{N_{4}}$, assuming a $\sord$ transition, together with a fit of the form
\eqref{eq:b4fss} where $N_{4}^{\omega}=\of{L^{pcr}}^{1/\nu}$. We also included higher order corrections. It can be seen that the fit seems to work fine, but the obtained values
$B_{4}^{cr}\fof{N_{2}}\,=\,-4.2\,\pm\,5.0$ and $\nu\,=\,2.01\,\pm\,0.74$ do not make much sense.}
  \label{fig:B4scale2}
\end{minipage}
\end{figure}

A more quantitative method to determine the order of a phase transition is to study finite-size scaling of the $\nord{4}$ Binder cumulant (Kurtosis) of the $N_{2}$ distribution,
$B_{4}\fof{N_{2}}=\frac{\avof{\of{N_{2}-\avof{N_{2}}}^{4}}}{\avof{\of{N_{2}-\avof{N_{2}}}^{2}}^{2}}$.
According to \cite{Binder}, this quantity should for large $N_{4}$ scale like
\[
B_{4}^{pcr}\fof{N_{2}}\of{N_{4}}\,\approx\,B_{4}^{cr}\fof{N_{2}}\,+\,c_{1}\,N_{4}^{-\omega},\label{eq:b4fss}
\]
where $B_{4}^{cr}\fof{N_{2}}$ is the critical, infinite volume value of the Binder cumulant. For a $\sord$ transition one should get $1<B_{4}^{cr}\fof{N_{2}}<3$ and $\omega\,=\,1/d_{H}\nu$, where
$\nu$ is the critical exponent of the correlation length $\xi_{N_{2}}\approx \abs{\kappa_{2}^{cr}-\kappa_{2}}^{-\nu}$ and $d_{H}$ the \emph{Hausdorff dimension}, whereas for a $\ford$ transition we should obtain $B_{4}^{cr}\fof{N_{2}}=1$ and
$\omega=1$.\\
We tried to fit our data for $B_{4}^{pcr}\fof{N_{2}}\of{N_{4}}$ assuming $1^{\text{st}}$ and $\sord$ scaling ansaetze (see Figs. \ref{fig:B4scale1} and \ref{fig:B4scale2}). As is typical for a weak
$\ford$ transition, the $\sord$ fit seems to work fine, but the obtained values $B_{4}^{cr}\fof{N_{2}}\,=\,-4.2\,\pm\,5.0$ and $\nu\,=\,1/d_{H}\omega\,=\,2.01\,\pm\,0.74$ do not make much sense.
Now, fixing $\omega=1$ for the $\ford$ ansatz, it is not possible to obtain $B_{4}^{cr}\fof{N_{2}}=1$ from a fit, not even when using only the data points from the largest two simulated systems:
in this case we obtain $B_{4}^{cr}\fof{N_{2}}\approx0.7$.
As the Binder cumulant involves a $\nord{4}$ moment, it requires rather large statistics which we probably have not accumulated yet for the largest system. A fit to the data of the next smaller pair
of systems, i.e. those consisting of 48k and 32k 4-simplices, leads to $B_{4}^{cr}\fof{N_{2}}\approx1.6$, which shows that our systems are presumably still too small and we will have to include
higher order scaling corrections to obtain a meaningful fit.\\

\section{Conclusion}
Our study confirms the qualitative findings of \cite{Bialas,deBakker}: for  $\kappa_{2}\,\approx\,\kappa_{2}^{pcr}\of{N_{4}}$ we find for $N_{4}\,\geq\,32$k a clear double peak structure in the
$N_{2}$ distribution, which becomes more pronounced with increasing system size (and there is no sign that the two peaks will eventually merge again in the thermodynamic limit). This is characteristic
of a weak $\ford$ transition. But the standard $\ford$ ansatz for finite-size scaling of $B_{4}^{pcr}\fof{N_{2}}$ did not work so far; presumably our systems are still too small and we have to include
higher order scaling corrections.

\end{document}